\documentclass[aps,pra,twocolumn,reprint,superscriptaddress]{revtex4}

\usepackage{hyperref}
\usepackage{graphicx}  
\usepackage{bm}        
\usepackage{amssymb}   
\usepackage{xspace}
\usepackage{verbatim}
\usepackage{color}
\usepackage{soul}
\usepackage{subfigure}
\usepackage[export]{adjustbox}
\usepackage{dcolumn}
\usepackage{amsthm,stmaryrd,mathtools} 
\usepackage{txfonts}
\usepackage{braket}
\usepackage{amsmath}
\usepackage{xcolor}
\usepackage{array}
\usepackage{multirow}
\usepackage{tabularx}
\usepackage{booktabs}
\usepackage{lipsum}

\begin{document}

\title{Long-range interacting Stark many-body probes with Super-Heisenberg precision}

\author{Rozhin Yousefjani}%
\email{rozhinyousefjani@uestc.edu.cn}
\affiliation{Institute of Fundamental and Frontier Sciences, University of Electronic Science and Technology of China, Chengdu 610051, China}

\author{Xingjian He}%
\affiliation{Institute of Fundamental and Frontier Sciences, University of Electronic Science and Technology of China, Chengdu 610051, China}

\author{Abolfazl Bayat}%
\email{abolfazl.bayat@uestc.edu.cn}
\affiliation{Institute of Fundamental and Frontier Sciences, University of Electronic Science and Technology of China, Chengdu 610051, China}

\begin{abstract}
In contrast to interferometry-based quantum sensing, where interparticle interaction is detrimental, quantum many-body probes exploit such interactions to achieve quantum-enhanced sensitivity. In most of the studied quantum many-body probes, the interaction is considered to be short-ranged. Here, we investigate the impact of long-range interaction at various filling factors on the performance of Stark quantum probes for measuring a small gradient field. These probes harness the ground state Stark localization phase transition which happens at an infinitesimal gradient field as the system size increases. Our results show that while super-Heisenberg precision is always achievable in all ranges of interaction, the long-range interacting Stark probe reveals two distinct behaviors. First, by algebraically increasing the range of interaction, the localization power enhances and thus the sensitivity of the probe decreases. Second, as the interaction range becomes close to a fully connected graph its effective localization power disappears and thus the sensitivity of the probe starts to enhance again. The super-Heisenberg precision is achievable throughout the extended phase until the transition point and remains valid even when the state preparation time is incorporated in the resource analysis. As the probe enters the localized phase, the sensitivity decreases and its performance becomes size-independent, following a universal behavior. In addition, our analysis shows that lower filling factors lead to better precision for measuring weak gradient fields. 
\end{abstract}

\maketitle


\section{Introduction}\label{I}
Quantum sensors can achieve unprecedented precision in measuring time~\cite{cacciapuoti2009space,ludlow2015optical}, electric~\cite{dolde2011electric,facon2016sensitive}, magnetic~\cite{budker2007optical,taylor2008high,tanaka2015proposed}, and gravitational fields~\cite{tino2019sage,aasi2013enhanced}, way beyond the capability of their classical counterparts. They can be manufactured in atomic scales and have found applications in a wide range of fields, from cosmology~\cite{Cosmology1, Cosmology2,xiong2021searching} to biology~\cite{Biology1, Biology2, shi2018single}. The precision of estimating an unknown parameter $h$, encoded in a quantum density matrix $\rho(h)$, is fundamentally bounded by Cram\'{e}r-Rao inequality as $\Delta h {\ge} 1{/}\sqrt{M\mathcal{F}}$, where $\Delta h$ is the standard deviation that quantifies the accuracy of the estimation, $M$ is the number of repetitions and $\mathcal{F}$ is a positive quantity called Fisher information. The scaling of Fisher information with respect to sensing resources, such as the probe size $L$, is a figure of merit that can be used for comparing the precision of different sensors. Typically, Fisher information scales algebraically with the size of the resource, namely  $\mathcal{F}{\propto}L^{\beta}$. 
In the absence of quantum features, classical sensing at best results in $\beta{=}1$, known as the standard limit. Quantum sensors, however, can achieve super-linear scaling with $\beta{>}1$ through exploiting quantum features such as entanglement~\cite{paris2009quantum,degen2017quantum}. 
Originally, enhancement in precision has been discovered for a special form of entangled states, known as GHZ states~\cite{greenberger1989going}, which results in $\beta{=}2$ also known as Heisenberg limit~\cite{giovannetti2004quantum,leibfried2004toward,boixo2007generalized,giovannetti2006quantum,banaszek2009quantum,giovannetti2011advances,frowis2011stable,wang2018entanglement,kwon2019nonclassicality}. 
Although there are several experimental demonstrations of GHZ-based quantum sensors~\cite{demkowicz2012elusive,albarelli2018restoring,GHZexp1, GHZexp2, GHZexp3}, their scalability is challenging due to the sensitivity of such delicate quantum states to decoherence. In addition, the interaction between particles in these probes is detrimental to their precision~\cite{de2013quantum,PhysRevA.90.022117,skotiniotis2015quantum}.

Strongly correlated many-body systems are very resourceful for realizing quantum technology tasks, such as sensing. These quantum probes, which harness the interaction between particles, are naturally scalable and expected to be more robust against decoherence. In particular, various forms of phase transitions in such systems have been used for achieving quantum-enhanced sensitivity, including first-order~\cite{raghunandan2018high,heugel2019quantum,yang2019engineering,ding2022enhanced}, second-order~\cite{zanardi2006ground,zanardi2007mixed,gu2008fidelity,zanardi2008quantum,invernizzi2008optimal,gu2010fidelity,gammelmark2011phase,skotiniotis2015quantum,rams2018limits,wei2019fidelity,chu2021dynamic,liu2021experimental,montenegro2021global,mirkhalaf2021criticality,di2023critical}, Floquet~\cite{mishra2021driving,mishra2022integrable}, dissipative~\cite{baumann2010dicke,baden2014realization,klinder2015dynamical,rodriguez2017probing,fitzpatrick2017observation,fink2017observation,ilias2022criticality}, time crystals~\cite{montenegro2023quantum, iemini2023floquet}, topological~\cite{budich2020non,sarkar2022free,koch2022quantum,yu2022experimental}, many-body~\cite{sahoo2023localization} and Stark localization~\cite{he2023stark} phase transitions. 
Other types of may-body probes profit from diverse measurement methods including    
adaptive~\cite{wiseman1995adaptive,armen2002adaptive,fujiwara2006strong,higgins2007entanglement,berry2009perform,said2011nanoscale,okamoto2012experimental,bonato2016optimized,okamoto2017experimental,fernandez2017quantum}, continuous~\cite{albarelli2017ultimate,gammelmark2014fisher,albarelli2018restoring,rossi2020noisy,yang2022efficient,ilias2022criticality}, and sequential~\cite{burgarth2015quantum,montenegro2022sequential} measurements.
Since most of the sensing proposals in many-body probes have been dedicated to short-range interactions, a key open problem  is whether long-range interactions can provide more benefits for sensing tasks?
Long-range interactions naturally arise in certain quantum devices, such as  ion traps~\cite{morong2021observation,smith2016many,rajabi2019dynamical} and
Rydberg atoms~\cite{choi2016exploring,rispoli2019quantum}. 
The nature of these interactions prevents the systematic study of their interesting physics and except for some models such as Lipshin-Meshkov-Glick (LMG)~\cite{chu2021dynamic,garbe2022critical}, and long-range Kitaev chain~\cite{Kitaev}, the effect of long-range interaction on  sensing precision remains almost untouched.    
\\

Gradient field sensing is of major importance in various fields, including biological imaging~\cite{waddington2020high,koonjoo2021boosting} and gravitometry~\cite{snadden1998measurement,griggs2017sensitive,stray2022quantum,phillips2022position}. In the former, the ultra-precise sensing of a weak gradient magnetic field increases imaging resolution, enabling the visualization of smaller tumors for early cancer detection. 
In the latter, precise gravity measurement is essential for  detection of gravitational waves~\cite{GravitionalWave1,GravitionalWave2}, investigating the equivalence principle~\cite{asenbaum2020atom}, obtaining the fine-structure~\cite{parker2018measurement} and measuring  Newton’s gravitational constant~\cite{rosi2014precision}.
Recently, we have shown that Stark probes can be exploited for measuring weak gradient fields with super-Heisenberg precision~\cite{he2023stark}, in which the scaling exponent $\beta$ can be as large as $\beta{\cong}6$. This sensor relies on Stark localization transition which could even happen in the presence of an infinitesimal gradient field in single- and multi-particle quantum systems.
The effect of a longer range of interaction on this sensor has not yet been explored. 
Addressing this issue is essential since the physical platforms for experimental realization of Stark localization, including ion traps~\cite{morong2021observation,smith2016many,rajabi2019dynamical} and Rydberg atoms~\cite{choi2016exploring,rispoli2019quantum} are naturally governed by long-range interactions.
\\

In this paper, we systematically study the effects of long-range interaction on the sensing capability of Stark probes.
We show that the strong super-Heisenberg scaling of the Stark probes persists even in the presence of long-range interaction and is achievable throughout  the extended phase of the system until the transition point.
Our results show that various range of interaction leaves distinct imprints on the scaling of the Fisher information. Making the interaction more long-ranged enhances the localization and, hence, decreases the value of the Fisher information and $\beta$. 
The localization effect disappears as the system gets closer to a fully connected graph and thus the sensitivity enhances again.
The achievable super-Heisenberg scaling remains valid even when the state preparation time is taken into account in resource analysis. 
Moreover, we provide a comprehensive investigation of the critical properties of long-range Stark probes and establish a concrete relationship between critical exponents of the system through an extensive finite-size scaling analysis.
We show that the enhanced sensitivity can be captured by measuring spin configurations in the relevant sector of the Hilbert space. 
Finally, we analyze the effect of filling factor (i.e., the number of excitations per site) on the sensing power of our Stark probes. While super-Heisenberg scaling is achievable for all studied filling factors, lower filling factors provide better precision.      
\\

This paper is organized as follows. 
We start by presenting the tools for assessing a quantum probe in section~\ref{II}.
After introducing our long-range Stark many-body probe in section~\ref{III}, we present the numerical results of sensing with the probe in the half-filling sector in section~\ref{IV}.
In the subsections of section~\ref{IV}, the scaling behavior of the probe, its critical properties, the resource analysis, and the optimal measurement  
are discussed.
Section~\ref{V} contains the analysis of the filling factor and the paper is summarized in  section~\ref{VI}.

\section{Ultimate precision limit}\label{II}
In this section, we briefly review the implications of Cram\'{e}r-Rao inequality for quantum sensing problems.  
In order to estimate an unknown parameter $h$ encoded in a probe, described by density matrix  $\rho(h)$,  one has to perform  a measurement which is described by a set of positive operator-valued measure  (POVM) $\{\Pi_{i}\}$. 
Each measurement outcome appears with the probability $p_i(h){=}\mathrm{Tr}[\Pi_{i}\rho(h)]$.
For this classical probability distribution one can show that the classical Fisher information (CFI), defined as~\cite{paris2009quantum,Rev1}
\begin{equation}\label{eq:CFI}
     \mathcal{F}_{C}(h)=\sum_i \frac{1}{p_{i}(h)}\left(\frac{\partial p_{i}(h)}{\partial h}\right)^{2},
 \end{equation}
establishes a bound on the estimation uncertainty $\Delta h$, known as Cram\'{e}r-Rao bound
\begin{equation}\label{eq:CCR}
    \Delta h{\ge}\frac{1}{\sqrt{M \mathcal{F}_C(h)}}.
\end{equation}
Where $M$ represnts the number of samples. The saturation of the inequality requires an optimal estimation algorithm. 
For large values of $M$, it is known that the optimal algorithm is the Bayesian estimator~\cite{Book1,Book2}.
Note that the CFI depends on the choice of measurement. 
In order to have a measurement-independent quantity, as an ultimate precision limit, one can maximize the CFI with respect to all possible measurements.
The corresponding quantity, known as Quantum Fisher Information (QFI), is thus defined as $\mathcal{F}_Q(h){=}\max_{\{\Pi_{i}\}}\mathcal{F}_C(h)$~\cite{meyer2021fisher,Rev2}. Therefore, the Cram\'{e}r-Rao inequality can be written in a hierarchical form as~\cite{giovannetti2011advances,Toth2014} 
\begin{equation}\label{eq:QCR}
    \Delta h{\ge}  \frac{1}{\sqrt{M \mathcal{F}_C(h)}} {\ge}\frac{1}{\sqrt{M \mathcal{F}_Q(h)}}.
\end{equation}
This hierarchical inequality shows that the QFI can be served as a benchmark for evaluating any sensing protocol with a given measurement setup through comparison of its corresponding CFI with the ultimate precision limit, quantified by the QFI $\mathcal{F}_Q$. 
Saturation of the ultimate precision bound, in Eq.~(\ref{eq:QCR}), relies on both selecting the optimal measurement basis and choosing the best estimation algorithm.
While the maximization with respect to measurement in the definition of the QFI seems notoriously challenging,  alternative methods can provide computational-friendly methods for calculating the QFI. 
In particular, it turns out that the QFI is related to a quantity called fidelity susceptibility $\chi(h)$ as $\mathcal{F}_Q{=}4\chi(h)$. 
The fidelity susceptibility is defined as~\cite{gu2008fidelity,gu2010fidelity,wei2019fidelity,FidelitySuscep3}
\begin{equation}\label{eq:FS}
\chi(h) = \dfrac{2 \big( 1 - \sqrt{\mathrm{Tr}[\rho(h)^{1/2}\rho(h{+}\delta h)\rho(h)^{1/2}]} \big )}{\delta h^2},  
\end{equation}
with $\delta h$ being an infinitesimal variation in $h$.
It has been shown that for systems that go through a  second-order quantum phase transition, the fidelity susceptibility and, hence, QFI show non-analytic behavior in the vicinity of the critical point~\cite{gu2008fidelity,FidelitySuscep1,FidelitySuscep2,
FidelitySuscep3,FidelitySuscep4}. 
This reflects the tremendous sensitivity of the system with respect to the control parameter $h$ which drives the system into the phase transition. 
In this paper, we rely on Eq.~(\ref{eq:FS}) for calculating the QFI and investigating the sensing power of a Stark many-body probe with long-range interaction.
\\

\section{Stark many-body probe}\label{III}

We consider a one-dimensional spin-1/2 chain of $L$ sites that is affected by a gradient field $h$.
While spin tunneling is restricted to nearest-neighbor sites, the interaction between particles is taken to be long-range  which algebraically decays by exponent $\eta{>}0$. The Hamiltonian reads
\begin{equation}\label{Eq.Hamiltonian}
H(h) = J\sum_{i=1}^{L-1}\left(\sigma_{i}^{x}\sigma_{i+1}^{x}+\sigma_{i}^{y}\sigma_{i+1}^{y}\right)+
\sum_{i<j}\dfrac{1}{|i-j|^{\eta}}
\sigma_{i}^{z}\sigma_{j}^{z} + h\sum_{i=1}^{L} i \sigma_{i}^{z},
\end{equation}
where $J$ is the exchange coupling, $\sigma_{i}^{(x,y,z)}$ are Pauli operators acting on site $i$, and $h$ is the amplitude of the applied gradient field, which has to be estimated. 
By varying the  power-law exponent $\eta$, one can smoothly interpolate between a fully connected graph ($\eta{=}0$) and a standard nearest-neighbor one-dimensional chain ($\eta{\rightarrow}\infty$).
Inherently, many interactions are long-range.
Coulomb and dipole-dipole interactions are notable examples of this interaction that can be modeled in certain quantum simulators, e.g., ion traps~\cite{morong2021observation,smith2016many,rajabi2019dynamical} and
Rydberg atoms~\cite{choi2016exploring,rispoli2019quantum}. 
The Hamiltonian Eq.~(\ref{Eq.Hamiltonian}) conserves the number of excitations in the $z$ direction, namely $[H,S_z]{=}0$, where $S_z{=}\frac{1}{2}\sum_i\sigma_i^{z}$. 
This implies that the Hamiltonian  is block-diagonal with respect to the number of excitations $N$. Hence, each block can be described by a filling factor of $n{=}N/L$. 
Here, we focus on the sensing power of our probe assuming that the filling factor $n$ is fixed and the probe is prepared in the lowest energy eigenstate of the relevant sector. 
Note that the true ground state of the Hamiltonian lies in the sector with $n{=}0$  (i.e., $N{=}0$ excitations). 
Nonetheless, throughout the paper, for the sake of convenience, we call the lowest eigenstate of the Hamiltonian for any given filling factor $n$ the ground state which should not be mistaken by the true ground state of the Hamiltonian at filling factor $n{=}0$.   

Regardless of the range of interaction, by increasing the strength of the field $h$, the probe
undergoes a quantum phase transition from an extended phase to a many-body localized one~\cite{kolovsky2008interplay,van2019bloch,schulz2019stark,yao2020many,chanda2020coexistence}. 
It is known that the many-body localization (MBL) transition occurs across the entire spectrum, in contrast to conventional quantum phase transition which occurs only at the ground state~\cite{gu2010fidelity}.
Detecting and characterizing the MBL transition across the whole spectrum usually rely on exact diagonalization  which severely restricts the numerical simulations to small systems~\cite{luitz2015many}. For analyzing the sensing power of a probe, one requires large system size behavior which is not accessible through exact diagonalization. Therefore, we exploit  Matrix Product State (MPS) simulation~\cite{tenpy}  to capture the behavior of QFI in large system sizes. 
While this allows us to extract a precise scaling analysis, it comes with the price that we will be limited to the ground state in each filling factor and cannot analyze the sensing power of excited states.

\section{Sensing at half-filling sector ($n{=}1/2$)}\label{IV}
\begin{figure}[t]
\includegraphics[width=\linewidth]{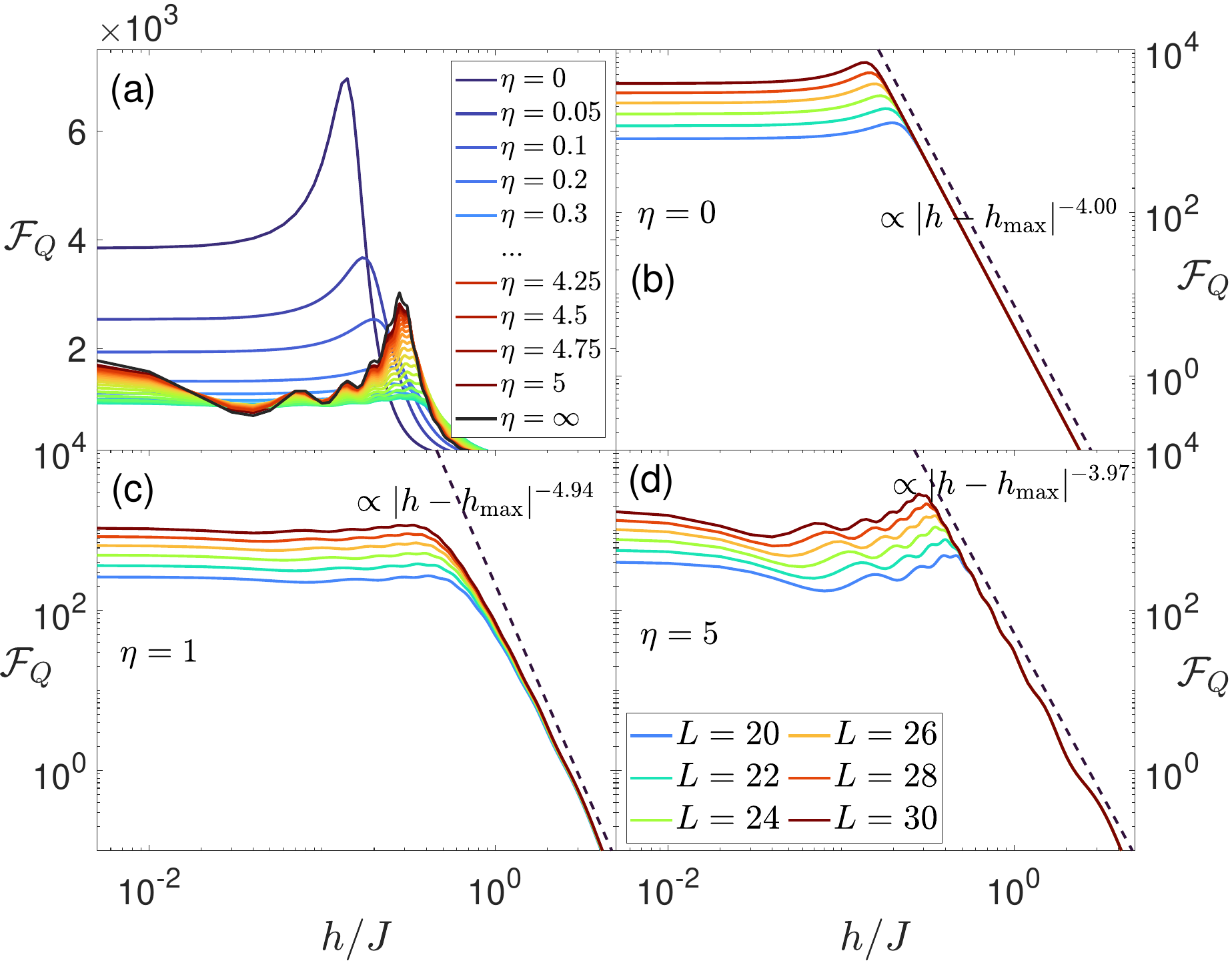}
\caption{(a) The QFI versus Stark field $h{/}J$ when our probe of size $L{=}30$ is prepared in the ground state of $H(h)$  with $n{=}1/2$ and different power-law exponents $\eta$.
(b)-(d) $\mathcal{F}_{Q}$ as a function of $h{/}J$ for probes of different sizes $L$ initialized in the ground state of $H(h)$ for $\eta{=}0,1$ and $5$, respectively.
The dashed lines in all panels are the best fit of $\mathcal{F}_{Q}$, namely $\mathcal{F}_{Q}{\propto}|h{-}h_{\max}|^{-\alpha}$, representing the size-independent algebraic behavior of the QFI in the localized phase.}\label{fig:Fig1}
\end{figure}

We first focus on the half-filling sector of the Hamiltonian in which we have $N{=}L/2$ excitations. 
In Fig.~\ref{fig:Fig1}(a), we plot $\mathcal{F}_{Q}$ as a function of Stark field $h{/}J$ for a probe of size $L{=}30$ with various choices of  $\eta$. Several interesting features can be observed.
First, by increasing $h{/}J$ the QFI shows a dramatic change in its behavior from being almost constant in the extended phase to a decreasing function in the localized regime.
During this transition, the QFI peaks at some $h_{\max}(\eta)$, 
which asymptotically converges to the transition point $h_{c}$ in the thermodynamic limit~\cite{FidelitySuscep3,FidelitySuscep4}. 
Second, various $\eta$'s leave distinct imprints on the QFI.
By moving from a fully connected probe ($\eta{=}0$) to a nearest-neighbor one ($\eta{\rightarrow}\infty$), the peaks of the QFI first decrease and then show a revival behavior.
This is because as $\eta$ decreases (i.e., interaction becomes more long-range) each spin configuration induces a different Zeeman energy splitting at any given site.
This effect is like random disorder potential, which helps the system to localize and thus reduces the QFI. 
The observed behavior continues until the system becomes close to a fully connected graph (for $\eta {\sim} 0.1$) in which all spin configurations induce almost the same energy splitting and thus the localization effect from off-resonant energy separations gradually disappears.  
Third, strong long-range interaction indeed enhances the sensitivity of the probe by  providing the highest value of $\mathcal{F}_{Q}$ in both the extended phase (i.e., $h{<}h_{\max}$) and at the transition point (i.e., $h{=}h_{\max}$).

To explore the behavior of the QFI in the thermodynamic limit, namely for $L{\rightarrow}\infty$, one can study the QFI for various system sizes. 
In Figs.~\ref{fig:Fig1}(b)-(d), we plot the ground state QFI as a function of Stark field $h{/}J$ for various system sizes $L$ and selected $\eta{=}0,1$ and $5$, respectively.
Regardless of the range of the interaction, by enlarging the probe size, the peak of the QFI increases and $h_{\max}$ gradually approaches zero, signaling the divergence of $\mathcal{F}_{Q}$ in the thermodynamic limit for a vanishing transition point $h_{c}{\rightarrow}0$.
While the finite-size effect can be seen in the extended phase, in the localized regime one deals with a size-independent algebraic decay of the QFI which can be perfectly described by $F_{Q}{\propto}|h{-}h_{\max}|^{-\alpha(\eta)}$ (dashed lines). 
From Figs.~\ref{fig:Fig1}(b)-(d), one can see that the exponent $\alpha$ takes the values $\alpha(\eta{=}0){=}4.00$, $\alpha(\eta{=}1){=}4.94$ and $\alpha(\eta{=}5){=}3.97$, respectively.

\subsection{Super-Heisenberg sensitivity}\label{IV. A}
\begin{figure}[t!]
\includegraphics[width=\linewidth]{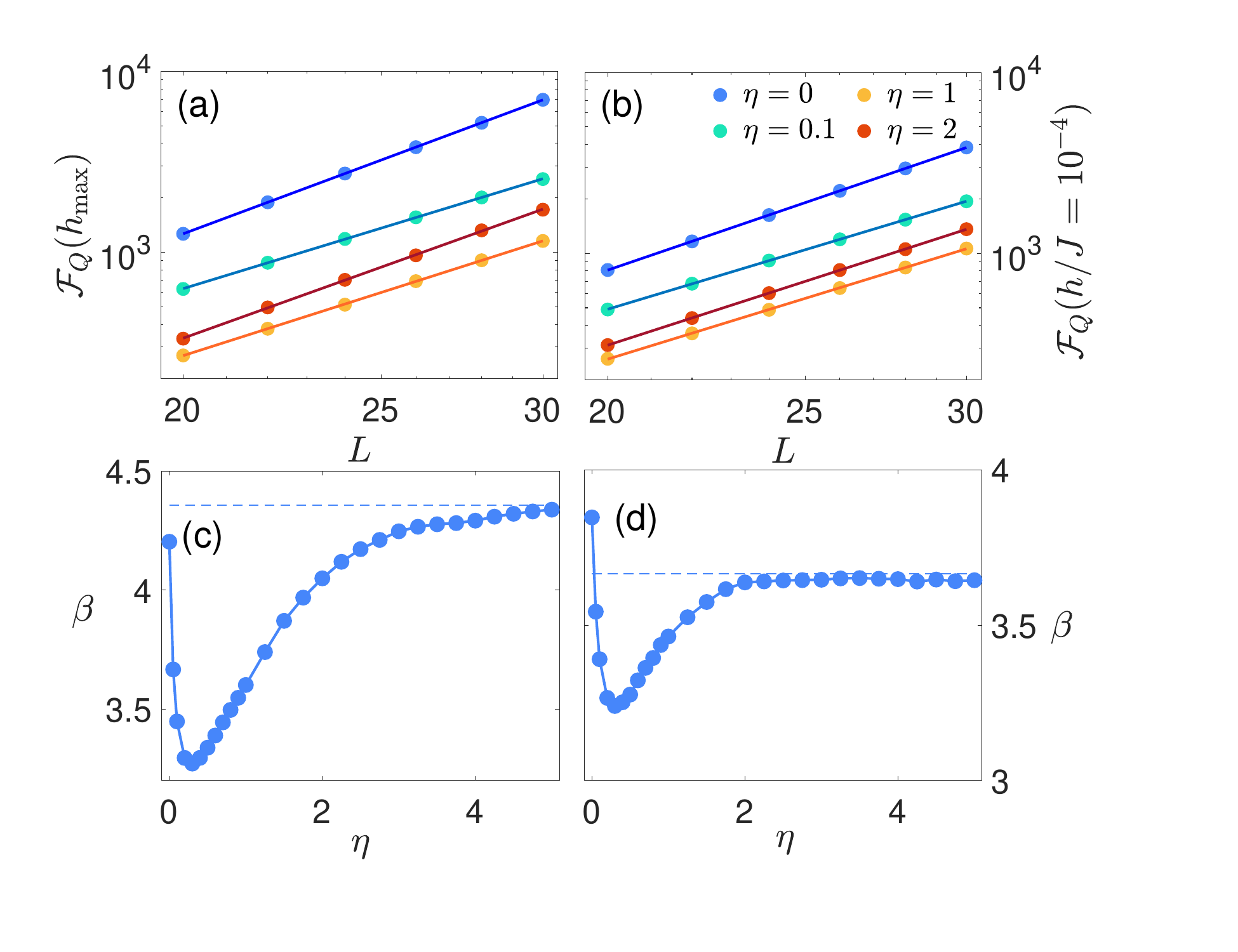}
\caption{Upper panels: the maximum of  QFI (markers) versus probe size $L$ for some values of $\eta$ in (a) transition point ($h{=}h_{\max}$) and (b) extended regime ($h{/}J{=}10^{-4}$). The lines are the best fitting function of the form $\mathcal{F}_{Q}{\propto}L^{\beta(h,\eta)}$. 
Lower panels: the scaling exponent $\beta(h,\eta)$ versus $\eta$ obtained (c) at the transition point and (d) in the extended phase.}\label{fig:Fig2}
\end{figure}
To characterize the scaling of the QFI with the probe size, in Figs.~\ref{fig:Fig2}(a) and (b), we plot $\mathcal{F}_{Q}$ versus $L$ for some values of $\eta$ both at the transition point, i.e., $h{=}h_{\max}$, and in the extended phase, i.e., $h{/}J{=}10^{-4}$, respectively.
In both panels, the markers represent the QFI obtained by numerical simulation and the lines are the best fitting function of the form $\mathcal{F}_{Q}(h,\eta){\propto}L^{\beta(h,\eta)}$.
The best obtained exponent $\beta(h,\eta)$ has been plotted as a function of $\eta$ in Figs.~\ref{fig:Fig2}(c) and (d), for $h{=}h_{\max}$ and  $h{/}J{=}10^{-4}$, respectively.
Some interesting observations can be highlighted.
First, regardless of the interaction range $\eta$, one can obtain super-Heisenberg sensitivity for our probe (i.e., $\beta{>}2$) both at the transition point and in the extended regime.
Second, as discussed before, by decreasing $\eta$ (i.e., making interaction more long-range) the effective Zeeman energy splitting enhances the localization and thus reduces the QFI as well as the exponent $\beta$. As $\eta$  further decreases, the probe becomes effectively fully connected, implying that all spin configurations induce equal energy splitting that does not contribute to the localization anymore. Therefore, $\beta$ changes its behavior and starts rising as $\eta$ decreases towards zero.

\subsection{Finite-size scaling analysis}\label{IV. B}
\begin{figure}[t!]
\includegraphics[width=\linewidth]{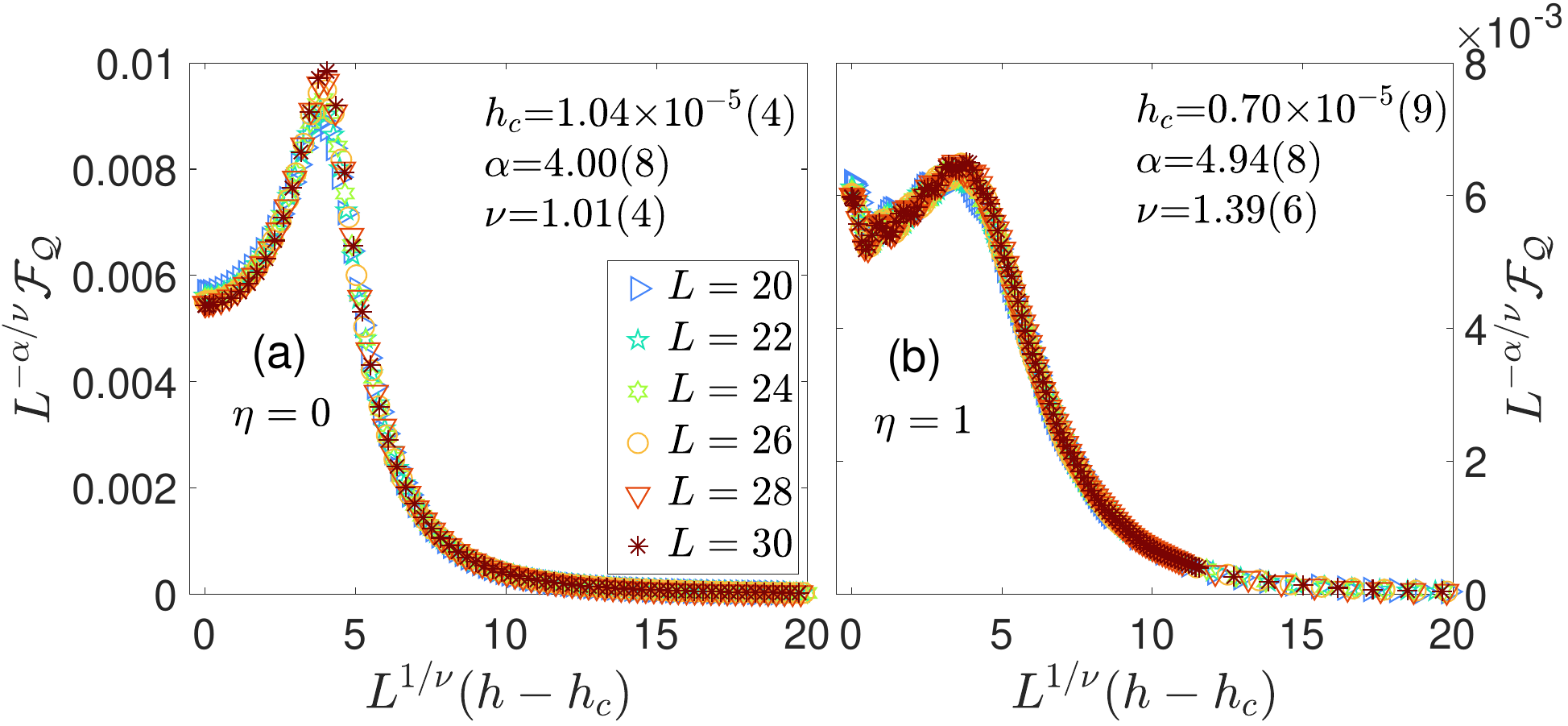}
\caption{Finite-size scaling analysis of the QFI following the ansatz Eq.~(\ref{Eq.finite-size-scalibg}) in (a) a fully connected probe $\eta{=}0$ and (b) a probe with $\eta{=}1$. The optimal data collapses are obtained for the attached critical properties $(h_{c}, \alpha, \nu)$ in each panel.  }\label{fig:Fig_FSA}
\end{figure}

The observed trend of the QFI in Figs.~\ref{fig:Fig1}(b)-(d)  (shown with dashed lines) strongly implies the algebraic divergence
of the QFI in the thermodynamic limit as $\mathcal{F}_{Q}{\propto}|h{-}h_{\max}|^{-\alpha}$.
For the sake of the abbreviation, we drop the dependency of the parameters on $\eta$ and $h$.   
This behavior which is attributed to all second-order phase transitions in the thermodynamic limit is accompanied by the emergence of a diverging length scale as $\xi{\sim}|h{-}h_c|^{-\nu}$, with $\nu$ known as the critical exponent.
To extract the parameters $\alpha$ and $\nu$ in finite-size systems one needs to establish finite-size scaling analysis.
In this technical method, the QFI is rescaled as 
\begin{equation}\label{Eq.finite-size-scalibg}
\mathcal{F}_{Q}=L^{\alpha/\nu}g(L^{1/\nu}(h-h_c)),
\end{equation}
where, $g(\cdot)$ is an arbitrary function.
Plotting the rescaled QFI, namely $L^{-\alpha/\nu}\mathcal{F}_{Q}$, versus $L^{1/\nu}(h{-}h_c)$ collapses all the curves of different probe sizes and the best data collapse can be obtained for accurate selection of critical properties, i.e., $(h_{c}, \alpha, \nu)$.
Figs.~\ref{fig:Fig_FSA}(a) and (b) illustrate the best-achieved data collapse for probes of size $L{=}20,\cdots,30$ for selected $\eta{=}0$, and $\eta{=}1$, respectively.
The critical properties for both panels, obtained using PYTHON package PYFSSA~\cite{melchert2009autoscalepy,andreas_sorge_2015_35293}, are  $(h_{c}, \alpha, \nu) {=}(1.04{\times} 10^{-5}, 4.00, 1.01)$, and $(h_{c}, \alpha, \nu) {=}(0.70{\times} 10^{-5}, 4.94, 1.39)$. 
For the sake of completeness, in Table~\ref{Tabel} we report the exponents $\alpha$ and $\nu$ for different values of $\eta$. 
Since in the finite-size systems, the peaks of the QFI at $h_{\max}$ are cutoff by the system size, one has $\mathcal{F}_{Q}{\propto}L^{\beta}$. 
The two expected behaviors of the QFI, namely $\mathcal{F}_Q{\propto}|h{-}h_c|^{-\alpha}$ in the thermodynamic limit and $\mathcal{F}_Q(h_{\max}){\propto}L^{\beta}$ for finite systems at the transition point, suggest a unified ansatz for the QFI as
\begin{equation}\label{Eq.ansatz}
\mathcal{F}_{Q}\propto\dfrac{1}{L^{-\beta} + A|h-h_{\max}|^{-\alpha}},
\end{equation}
where $A$ is a constant. One can indeed retrieve the two behaviors from the above ansatz by either choosing $L{\rightarrow} \infty$ or $h{=}h_{\max}$.
Note that, the two ansatzes of Eqs.~(\ref{Eq.finite-size-scalibg}) and (\ref{Eq.ansatz}) describe the same quantity and thus have to match with each other. A simple factorization of $L^{-\beta}$ from the denominator of Eq.~(\ref{Eq.ansatz}) shows that the two ansatzes are the same provided that the exponents satisfy
\begin{equation}\label{Eq.parameters}
\beta = \frac{\alpha}{\nu}.
\end{equation}
The validity of the above equation for all the considered $\eta$'s is evidenced in the presented data in Table~\ref{Tabel} in which 
$\alpha/\nu$, obtained from finite-size scaling analysis of Eq.~(\ref{Eq.finite-size-scalibg}), matches closely with $\beta$, obtained from scaling analysis in Fig.~\ref{fig:Fig2}(a).
\begin{table*}[t!]
\resizebox{\linewidth}{!}{
\setlength{\tabcolsep}{1mm}{
\renewcommand{\arraystretch}{1.7}
\begin{tabular}{|c|c|c|c|c|c|c|c|c|c|c|c|c|c|c|c|c|c|c|c|c|c|c|c|c|c|c|c|c|}
\specialrule{0em}{3pt}{3pt}
\hline
$\eta$   & 0    & 0.05 & 0.10  & 0.20  & 0.30  & 0.40  & 0.50  & 0.60  & 0.70  & 0.80  & 0.90  & 1.00    & 1.25 & 1.50  & 1.75 & 2.00    & 2.25 & 2.50  & 2.75 & 3.00    & 3.25 & 3.50  & 3.75 & 4.00    & 4.25 & 4.50  & 4.75 & 5.00    \\ \hline
$\alpha$   & 4.00 & 4.18 & 4.59 & 5.01 & 5.24 & 5.33 & 5.32 & 5.27 & 5.20 & 5.11 & 5.02 & 4.94 & 4.74 & 4.59 & 4.46 & 4.35 & 4.27 & 4.21 & 4.15 & 4.11 & 4.08 & 4.05 & 4.03 & 4.01 & 4.00 & 3.99 & 3.98 & 3.97 \\ \hline
$\nu$   & 1.01 & 1.18 & 1.35 & 1.53 & 1.61 & 1.63 & 1.61 & 1.57 & 1.53 & 1.48 & 1.43 & 1.39 & 1.28 & 1.22 & 1.17 & 1.12 & 1.10  & 1.08 & 1.06 & 1.04 & 1.03 & 1.02 & 1.02 & 1.01 & 1.01 & 1.01 & 1.00    & 1.00    \\ \hline
$\alpha/\nu$ & 3.96 & 3.54 & 3.40 & 3.27 & 3.25 & 3.27 & 3.31 & 3.36 & 3.40 & 3.45 & 3.51 & 3.55 & 3.71 & 3.76 & 3.81 & 3.89 & 3.88 & 3.89 & 3.91 & 3.95 & 3.96 & 3.97 & 3.95 & 3.97 & 3.96 & 3.95 & 3.98 & 3.97 \\ \hline
$\beta$   & 4.20 & 3.67 & 3.45 & 3.30 & 3.27 & 3.30 & 3.34 & 3.39 & 3.44 & 3.50 & 3.55 & 3.60 & 3.74 & 3.87 & 3.97 & 4.05 & 4.12 & 4.17 & 4.21 & 4.25 & 4.27 & 4.28 & 4.28 & 4.29 & 4.31 & 4.32 & 4.33 & 4.34 \\ \hline
\end{tabular}
}
}
\caption{Extracted critical exponents including $\alpha$, $\nu$, their ratio $\alpha/\nu$, and the exponent $\beta$ for various values of $\eta$. Here, reported $\alpha$ and $\nu$ that control the speed of algebraic divergence of the QFI $\mathcal{F}_{Q}{\propto}|h{-}h_{\max}|^{-\alpha}$ and the  length scale $\xi{\sim}|h{-}h_c|^{-\nu}$ in the thermodynamic limit, respectively, are obtained through finite-size scaling analysis. As it is evident from the values in the table, $\alpha{/}\nu$  and $\beta$ are very close to each other, which guarantees the validation of Eq.~(\ref{Eq.parameters}) for all ranges of interaction. Small deviations between $\alpha{/}\nu$  and $\beta$ are due to finite-size effects.}
\label{Tabel}
\end{table*}

\subsection{Resource analysis}\label{IV. C}
\begin{figure}[t]
\includegraphics[width=\linewidth]{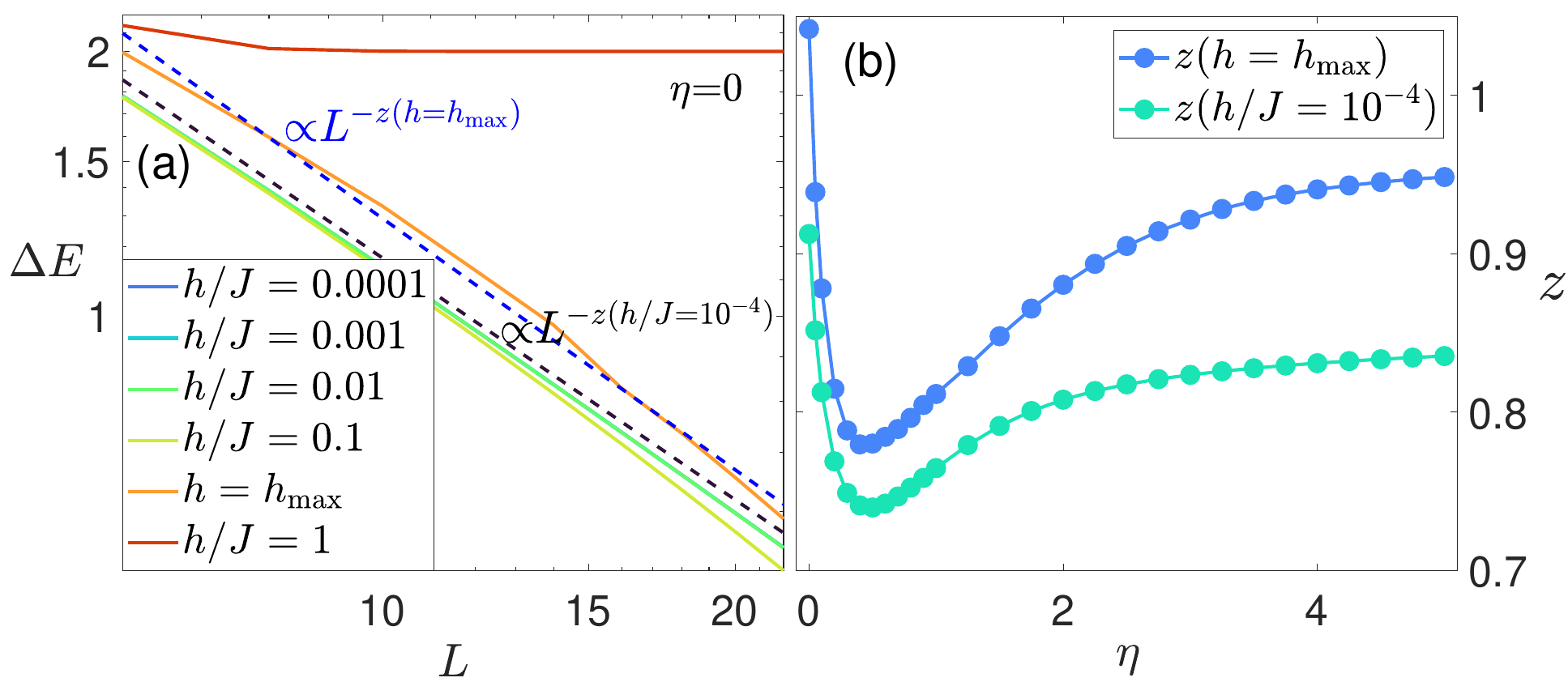}
\caption{(a) The energy gap $\Delta E$ as a function of the probe size $L$ in a fully connected probe (i.e., $\eta{=}0$) for different values of Stark field $h{/}J$. The results are obtained using exact diagonalization. The curves for $h{/}J{=}0.0001$ to $h{/}J{=}0.01$, in the extended phase, overlap with each other. The dashed lines are the best fitting function as $\Delta E{\propto} L^{-z}$ with $z(h{/}J{=}10^{-4}){\simeq}0.91$ and $z(h{=}h_{\max}){\simeq}1.04$ for the probe in the extended phase and at the transition point, respectively.
(b) Extracted dynamical exponent $z$ for all ranges of the interaction (i.e., $\eta$) in both the extended phase ($h/J{=}10^{-4}$), and at the transition point ($h{=}h_{\max}$).}\label{fig:Fig_zeta}
\end{figure}

Up to now, we showed that quantum criticality can indeed offer significant advantages for quantum sensing.
Nevertheless, this advantage is usually hindered by the time required to prepare the ground state close to the critical points.
Initializing a probe in its ground state via, for instance, adiabatic evolution~\cite{liu2021experimental},  demands a time that scales with the probe size as $t{\propto}L^{z}$~\cite{rams2018limits}, in which the exponent $z$ is known as dynamical exponent and determines the rate of the energy gap closing, namely $\Delta E{\propto}L^{-z}$, for  a system approaching to its criticality. 
Taking initialization time into consideration offers the normalized QFI, i.e., $\mathcal{F}_{Q}{/}t$ as a new figure of merit~\cite{rams2018limits,chu2021dynamic,montenegro2022sequential}. 
Since $\mathcal{F}_{Q}(h_{\max}){\propto} L^{\beta}$ one can easily show that the normalized QFI scales as 
\begin{equation}
    \mathcal{F}_Q{/}t\propto L^{\beta-z}. 
\end{equation}
In order to estimate the dynamical exponent $z$, one has to numerically compute the energy gap $\Delta E$ versus the system size $L$.  
In Fig.~\ref{fig:Fig_zeta}(a), we plot energy gap $\Delta E$ obtained through exact diagonalization as a function of $L$ for a fully connected probe ($\eta{=}0$) in the extended phase (i.e., $0.0001{\leqslant}h{\leqslant}0.1$), at the transition point (i.e., $h{=}h_{\max}$) and in the localized phase (i.e., $h{/}J{=}1$). 
An algebraic decay as a function of $L$ for energy gap is observed in the extended phase, with $z{=}0.91$, at the transition point, with $z{=}1.04$, and in the localized phase, with $z{=}0$. 
In Fig.~\ref{fig:Fig_zeta}(b), we plot the dynamical exponent $z$ as a function of $\eta$ for a probe in the extended phase ($h{/}J{=}10^{-4}$) and at the transition point ($h{=}h_{\max}$). 
As the results show, the exponent $z$ qualitatively behaves similarly to the exponent $\beta$ as the interaction range $\eta$ varies. 
It is worth emphasizing that even by considering time into the resource analysis, the exponent $\beta{-}z$ remains larger than 2 in all interaction ranges. This super-Heisenberg scaling can indeed provide a significant advantage for weak-field sensing.

\begin{figure}[t!]
\includegraphics[width=0.65\linewidth]{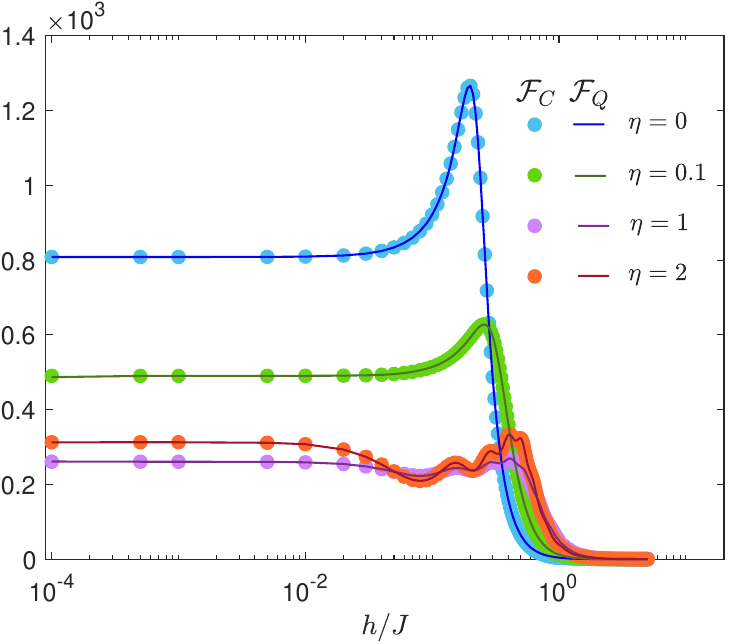}
\caption{ The QFI (filled lines) and the CFI (markers) versus Stark filed $h/J$ for a probe of size $L=20$ prepared in the ground state of Hamiltonian Eq.~(\ref{Eq.Hamiltonian}) with different $\eta$'s.}\label{fig:OM}
\end{figure}

\subsection{Optimal measurement}\label{IV. D}
Recovering the precision enhancement
offered by the QFI, generally, demand performing complex measurement that may dependent on
the unknown parameter. 
This makes it crucial to provide a suboptimal
yet local and experimentally achievable set of measurements
that capture the precision enhancement.
Remarkably, in our probe measuring spin configurations described by observable $\Pi_{i=1}^{L}\sigma_{i}^{z}$ in the sector $S_{z}{=}0$ closely saturates the  Cram\'{e}r-Rao inequality Eq.~(\ref{eq:QCR}). 
To show this, in Fig.~\ref{fig:OM} we plot both CFI (markers) and QFI (lines)  captured by Eqs.~(\ref{eq:CFI}) and (\ref{eq:FS}), respectively.
The curves show the sensitivity of the probe as a function of the Stark field $h$ for a system of size $L{=}20$ which is prepared in the ground state of $H(h)$ with  $\eta{\in}\{0,0.1,1,2\}$. 
Clearly, regardless of the interaction range, the CFI and QFI closely match showing that spin configuration is indeed an optimal measurement. The maximum of CFI happens exactly at the transition point, namely $h_{max}$, and this quantity resemble the behavior of QFI respect to $\eta$.

\section{Filling factor analysis}\label{V} 
\begin{figure}[t!]
\includegraphics[width=\linewidth]{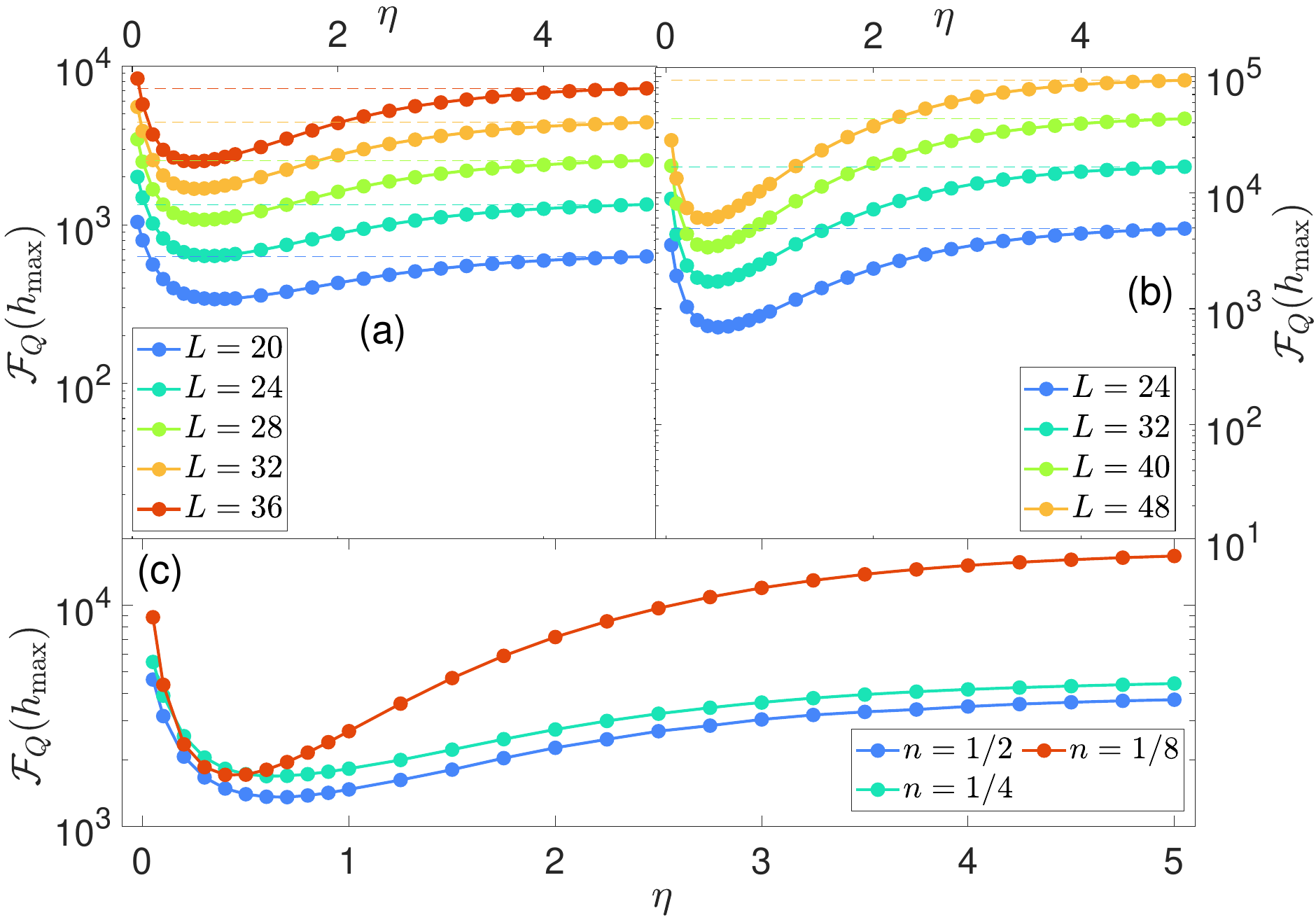}
\caption{The maximum  of the ground state QFI at the transition point $h{=}h_{\max}$ versus the power-law exponent $\eta$ for filling factors of: (a) $n{=}1/4$; and (b) $n{=}1/8$.
(c) The achievable maximum of the QFI at the transition point as a function of $\eta$ in a probe of size $L{=}32$, initialized in sectors of different filling factors of $n{=}1/2, 1/4$ and $1/8$.  }\label{fig:QFI_FF}
\end{figure} 
Having described the many-body Stark probe in a half-filling sector of the Hilbert space, we now focus on the effect of the filling factor $n$ on the performance of our sensor.
In Figs.~\ref{fig:QFI_FF}(a) and (b) we plot the QFI at the transition point $h{=}h_{\max}$ as a function of $\eta$ for filling factors $n{=}1/4$ and $n{=}1/8$, respectively. 
Clearly, analogs to the scenario of $n{=}1/2$ (see Fig.~\ref{fig:Fig1}(a)) as $\eta$ decreases (the interaction becomes more long-range) the QFI goes down and then revives as the effective localization impact disappears.
Interestingly, for larger filling factors (e.g. $n{=}1/2$ and somehow $n{=}1/4$), a fully connected probe with $\eta{=}0$ outperforms the other choices of $\eta$. As the filling factor reduces, the best performance belongs to the nearest-neighbor probe with $\eta{\rightarrow}\infty$.
In addition, our results evidence that decreasing $n$ can remarkably boost the achievable QFI.
This can be observed in Fig.~\ref{fig:QFI_FF}(c) which represents $\mathcal{F}_{Q}(h_{\max})$ in a probe of size $L{=}32$ prepared in various sectors of $n{=}1/2, 1/4$ and $1/8$.
These results are in line with our previous results in which the highest advance was obtained for a Stark probe with single excitation~\cite{he2023stark}.
 
\begin{figure}[t!]
\includegraphics[width=\linewidth]{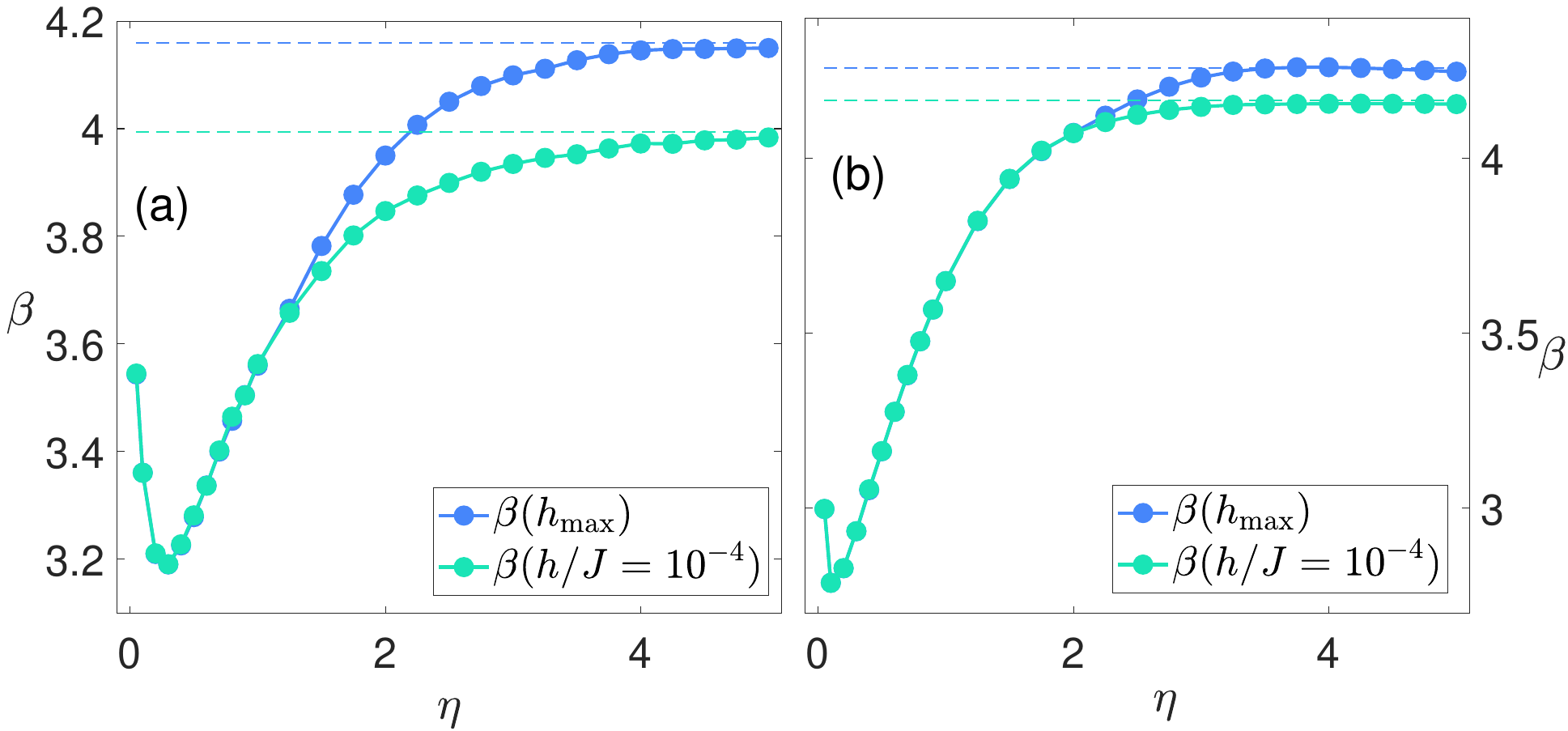}
\caption{The scaling of the QFI, namely $\mathcal{F}_{Q}(h,\eta){\propto}L^{\beta(h,\eta)}$, as a function of $\eta$ for (a) $n{=}1/4$ and (b) $n{=}1/8$. In both panels the extracted $\beta$ is reported for the transition point $h{=}h_{\max}$ and the extended phase $h{/}J{=}10^{-4}$. }\label{fig:Beta_FF}
\end{figure}

To characterize the impact of the filling factor on the scaling of the QFI with respect to $L$, similar to the scenario of the $n{=}1/2$, we fit the obtained QFI for different probe size $L$ with function $\mathcal{F}_{Q}{\propto}L^{\beta(h,\eta)}$. 
The best fits result in reported $\beta$'s as a function of $\eta$ in Figs.~\ref{fig:Beta_FF}(a) and (b) for $n{=}1/4$ and $n{=}1/8$, respectively. 
In each panel, we report the obtained $\beta$ at the transition point ($h{=}h_{\max}$) as well as in the extended phase ($h{/}J{=}10^{-4}$).
As the Figs.~\ref{fig:Beta_FF}(a) and (b) show, the exponent $\beta$ shows qualitatively similar behavior to the half-filling case as the interaction becomes more long-ranged. Importantly, for all interaction ranges the exponent $\beta$ shows super-Heisenberg scaling, and the best performance is always obtained for a nearest-neighbor probe. 
By decreasing the filling factor $n$, the performance of the probe in the extended phase gets closer to the one at the transition point.
This is in full agreement with our previous results obtained for the Stark probe with single particle~\cite{he2023stark} in which for the nearest-neighbor probe both cases yield the same $\beta$.

\section{Conclusion}\label{VI}
Stark localization transition in many-body systems, as a result of applying a gradient field in the lattice, has been harnessed to generate an ultra-precise sensor for measuring weak gradient fields. 
In this paper, we addressed the effect of long-range interactions on the capability of these probes. 
Our study showed that strong super-Heisenberg precision of the Stark probe can be obtained in all ranges of interaction in the extended  phase until the transition point. However, as the interaction becomes more long-range two different behaviors can be observed.
Initially, by making the system more long-ranged the sensing power, quantified by QFI and its exponent $\beta$, decreases. 
Then, around $\eta{\sim} 0.1$, where the system becomes effectively a fully connected graph, the sensitivity enhances again which can be seen in the rise of both QFI and $\beta$.
These different trends can be explained through long-range interaction induced localization. 
In long-range interacting systems, keeping the filling factor fixed, every given spin configuration induces a different Zeeman energy splitting at each site. 
This energy splitting behaves like an effective random disorder that enhances localization and decreases the sensing power.  
When the interaction becomes almost fully connected, the energy splitting of all spin configurations becomes equal and effective localization disappears, which boosts the sensitivity of the probe.
Interestingly, even by incorporating state preparation time in our resource analysis, the super-Heisenberg scaling still remains valid. 
In the localized phase, the system becomes size-independent and QFI follows a universal function. 
Several critical exponents governing the localization transition as well as their relationship have been extracted through extensive finite-size scaling analysis. 
In addition, we show that the CFI obtained via measuring  spin configurations closely matches with the QFI, showing that spin configuration is indeed an optimal measurement.
Finally, we have shown that the sensitivity decreases by increasing the filling factor.

\section*{Acknowledgment}
A.B. acknowledges support from the National Key R\&D Program of China (Grant No. 2018YFA0306703), the National Science Foundation of China (Grants No. 12050410253, No. 92065115, and No. 12274059), and the Ministry of Science and Technology of China (Grant No. QNJ2021167001L). R.Y. thanks the National Science Foundation of China for the International Young Scientists Fund (Grant No. 12250410242).


\end{document}